# A Mixed-Methods Study of Classroom Community Among Undergraduate Students in Introductory Mathematics and Statistics

Shira Viel[1], Maria Tackett[2], Sarwari Das[3], and Joseph Choo[2]
[1]Department of Mathematics, Duke University
[2]Department of Statistical Science, Duke University
[3] Master in Interdisciplinary Data Science, Duke University

## Abstract

A strong sense of classroom community is associated with many positive learning outcomes and is a crucial contributor to undergraduate students' persistence in STEM, particularly for women and students of color. This article describes a mixed-methods investigation into the relationship between classroom community and course attributes in introductory undergraduate mathematics and statistics courses, mediated by student identities. The project was originally motivated by and conducted amid the Covid-19 pandemic: data were collected from online courses in the 2021-21 academic year and from hybrid and in-person courses in Fall 2021. Quantitative data were gathered from both students and instructors and analyzed using structural equation modeling. The primary survey instrument is the validated Classroom Community Scale – Short Form. These quantitative results are complemented and contextualized by thematic and textual analyses of focus group data, gathered using a newly developed protocol piloted during the 2021-22 academic year. All data come from a private university in the United States. Preliminary practical implications of the study include the value of synchronous participation in fostering connectedness and the importance of attending to students' personal identities in understanding their experiences of belonging.

# Introduction

Students' sense of belonging, particularly in introductory mathematics courses, plays a critical role in their persistence in STEM (Ellis et al., 2014; Olson & Riordan, 2012; Seymour, 1997; Seymour et al., 2019; Tinto, 1993). More broadly, students' sense of classroom community, which as defined by Rovai encompasses both a sense of belonging as well as a shared commitment to learning (2002a), has been shown to have associations with many positive learning qualities. These include academic motivation and achievement (Freeman et al., 2007; McKinney et al., 2006; Summers & Svinicki, 2007), perceived learning (Rovai 2002b; Trespalacios & Perkins, 2016), and knowledge-sharing behaviors and academic self-efficacy (Yilmaz, 2017). Classroom community is particularly important in courses with an online component, where feelings of disconnectedness may arise due to physical distancing (Green et al., 2017; Richardson et al., 2017). Indeed, the literature shows that students in face-to-face courses consistently feel a stronger sense of community than do those in remote courses (see, e.g., Vavala et al., 2010).

Given the well-established connections between students' sense of classroom community and positive learning behaviors and outcomes, it is critical for practitioners to understand how their pedagogical practices and policies impact classroom community. Understanding the way these associations may be mediated by the diverse aspects of student identities is particularly important, as it is known that experiences of classroom community can differ significantly for students from different backgrounds (see, e.g., Rovai & Ponton 2019), and that feeling connected and supported in the STEM classroom is especially important for the persistence of women and students of color (Seymour et al., 2019). Given the pivotal role that introductory undergraduate courses in mathematics and statistics play in STEM persistence (see, e.g., Ellis et al., 2014;

Seymour, 1997; Seymour et al., 2019), studying classroom community in this setting is particularly critical. This article describes a mixed-methods project begun in Summer 2020, motivated by both the Covid-19 pandemic and the racial justice movement. The time was ripe to investigate how classroom practices impact student sense of belonging amidst the challenges of remote learning and systemic racial inequality. We sought to answer the following research question:

> *How is undergraduate students' sense of classroom community in introductory mathematics and statistics affected by course attributes and student identities?*

## Theory and Background

**Classroom Community**

We take our theory of classroom community from Alfred Rovai, who in turn draws heavily upon the theory of community developed by McMillan & Chavis. The latter authors propose that a sense of community is comprised of feelings of belonging and mattering as well as a "shared faith that members' needs will be met through their commitment to be together" (McMillan & Chavis, 1986). Rovai specifies this definition to the classroom setting as follows:

> *One can define classroom community as a feeling that members have of belonging, a feeling that members matter to one another and to the group, that they have duties and obligations to each other and to the school, and that they possess shared expectations that members' educational needs will be met through their commitment to shared learning goals* (2002b, p.322).

More explicitly, Rovai identifies two key components of classroom community:

1. Connectedness, or "the feeling of belonging and acceptance" (2002b, p.322), and

2. Learning (support), or "commonality of learning expectations and goals" (2002b, p.322).

We follow Rovai in using the terms "connectedness" and "belonging" interchangeably: this aspect of classroom community is precisely what Seymour et al. showed to be an important factor in persistence in STEM (2019). We do note, however, that some say that "belonging" has a more subjective connotation due to its association with identity (Crisp, 2010). We also note that Rovai uses the singular "learning" to refer to the second component of classroom community, but we use here the term "learning support" to emphasize that the component does not refer to actual learning outcomes, but rather to students' perception of their common learning goals. Others have also identified this potential source of confusion in terminology: we follow Demmans Epp, et al in using the term "learning support" (2020); another tack is to refer to the two components as "social community" and "learning community" as in (Dawson, 2008) and (Rovai and Ponton, 2009).

**The Classroom Community Scale**

Rovai's 20-item Classroom Community Scale, or CCS, was developed in 2002, with two interpretable factors corresponding to the two key components of classroom community defined above, connectedness and learning (support) (Rovai, 2002a).

The CCS was initially validated for graduate students in online courses (Rovai 2002a). While this population remains a popular setting for the instrument (see, e.g., Ahmady et al., 2018; Beeson et al., 2019; Rovai, 2002b; Trespalacios & Perkins, 2016; Wiest, 2015), the CCS is now used widely to assess classroom community among postsecondary students enrolled in courses of all formats, including undergraduates enrolled in online courses (Kocdar et al., 2018; Robinson et al., 2019; Young & Bruce, 2011), undergraduates enrolled in face-to-face courses (Brown & Pederson, 2020; Kirby & Thomas 2021), and undergraduates enrolled in courses with both in-person and online components (Dawson 2008). In addition, the CCS has been used to measure differences in senses of classroom community between undergraduates studying remotely and those studying in-person (Parrish et al 2023; Rovai & Jordan, 2004; Vavala et al., 2010). There are two other instruments in addition to the CCS that appear often in the context of measuring classroom community in online undergraduate courses (Zimmerman & Nimon, 2017; Dellasega 2021): the Community of Inquiry Scale (Arbaugh et al., 2008) and the Online Student Connectedness Survey (Bollinger & Inan, 2012). However, the CCS is likely the most frequently used of these three (Randoph & Crawford 2013), particularly in face-to-face environments: efforts to validate the other two scales for offline courses are very recent (Ariati et al 2023).

Cho & Demmans Epp developed the 8-item Classroom Community Scale Short Form (CCS-SF) in 2019 to both address criticisms of low construct validity in the original CCS (Barnard-Brak & Shiu, 2010), and to make the already relatively short instrument even shorter, to reduce respondent fatigue and increase the likelihood of completion (2019). The instrument was initially validated with a population of online students enrolled in Massive Open Online Courses (MOOCs) and graduate courses (Cho & Demmans Epp, 2019). Discussion of the CCS-SF has been sparse (Akatsuka, 2020; Demmans Epp et al., 2020), with no efforts to further validate the instrument until the authors of this article undertook to do so. We succeeded in validating the CCS-SF for undergraduate students in online courses (in introductory mathematics and statistics), and then performed confirmatory analysis to validate it for undergraduate students taking in-person courses (Tackett et al., 2022).

**Study Context and Overview**

Prior work studying classroom community among undergraduates has established that sense of community is higher in face-to-face courses (see, e.g., Vavalva et al., 2012), and in courses employing active learning techniques (see, e.g., Brown and Pederson, 2020), including Team Based Learning (Parrish et al., 2021). However, none of these studies have focused specifically on introductory quantitative science courses, which tend to be more lecture-based. In the realm of mathematics, prior work has tended to focus only on students' sense of belonging (see, e.g., Lahdenperä & Nieminen, 2020; Good et al., 2012), rather than the broader sense of classroom community, which also encompasses support for and shared goals of learning.

This study aims to fill these gaps, analyzing how introductory undergraduate mathematics and statistics students' sense of classroom community is impacted by course attributes, such as synchronous delivery and presence of groupwork, and how these impacts may be mediated by aspects of student identities, such as gender and race.

We take a mixed-methods approach to our investigation to obtain a deeper and more comprehensive understanding of the research question than either method alone (Cresswell, 2009; Dawadi et al., 2021). The primary instrument for data collection was the Classroom Community Scale – Short Form (Cho & Demmans Epp, 2019). The resulting quantitative data were analyzed using structural equation modeling. A focus group protocol was then developed and piloted to collect qualitative data, which were analyzed using thematic and textual methods to enrich and contextualize our quantitative results.

## Quantitative Analyses

**Methodology**

*Instruments*

Student data were collected through a survey comprised of the 8-item Classroom Community Scale – Short Form (Cho & Demmans Epp, 2019), 4 demographic questions (class year, gender, race, and disability status), 2 questions on math experience (prior course in the subject at the institution and grade received), and 2-5 questions, depending on the semester, on how the student was taking classes (Number of classes enrolled, living on campus or off, mode of class delivery, mode of class participation). Course-level data on mode of delivery and classroom practices were collected via a brief instructor interview or survey.

*Population*

Data come from undergraduate students and instructors in introductory mathematics and statistics classes at a private university in the United States. For the purposes of this study, "introductory" is defined as a course with no prerequisites or a course that students can take based solely on high school credentials, such as Advanced Placement (AP) exam results. The entry-level mathematics courses at the university are Calculus I and Calculus II, along with a two-semester Calculus I course sequence that incorporates precalculus content. The introductory statistics classes include introductory statistics and two introductory data science courses within the statistics department. Data collection took place for three semesters, from Fall 2020 through Fall 2021. All students over the age of 18 enrolled in one of these courses during this time frame were eligible to participate in the survey.

*Data collection*

Surveys were administered by the respective course instructors during the last month of the semester. Survey completion was voluntary, and students who completed the survey by the requested deadline were entered into a drawing for two $25 Amazon gift cards. Course instructors completed either a brief interview or a short survey about the course structure and classroom practices. The student and course data were combined and de-identified prior to

analysis, so no student or course could be directly identified in the data set or results. The data were collected and analyzed under IRB Protocol # 2021-0050.

*Methods of Analysis*

The survey data were first summarized using visualizations and descriptive statistics. Then, Structural Equation Models (SEM) were used to examine associations between classroom practices, student identity, and the two factors of student sense of community. SEMs, which are frequently used in social sciences and education research such as this study, were employed here because they made it feasible to model the two latent classroom community factors of connectedness and learning support. (These factors were not asked about directly in the student survey.) A threshold of 0.05 was used to determine which variables had a significant effect on the latent response variables, where each latent response variable corresponded to one of the two latent factors of student sense of community. All analysis was conducted in R; the *tidyverse* package (Wickham et al, 2019) was used for descriptive statistics and visualizations and the *lavaan* package (Rosseel, 2012) was used for the SEMs.

Each latent variable was created using the responses to the survey items corresponding to the respective factor of connectedness or learning support. Previous research has shown an association between students' sense of connectedness and their perception of learning environment and ultimately their perceived learning and achievement outcomes (Rovai 2002b; Trespalacios & Perkins, 2016). Thus, the models were constructed such that connectedness was one of the predictors of learning support.

Variables related to course structure, student identity, and student educational activities within the course were all considered as potential predictors in the models. These predictor variables are described in more detail below.

*Course structure variables*

The course structure variables were primarily related to the classroom experience and course characteristics that might encourage student interaction. Many course structure variables were considered, but here we present the primary ones of interest based on preliminary analysis results. The first type of course structure variable is related to the presence of collaborative student work. This aspect was captured by three variables: **work groups** (indicator of whether at least 50% of synchronous class time was used for collaborative work), **project groups** (indicator of whether students worked in the same group for multiple weeks)**,** and **graded groups** (indicator of whether group work was graded)**.**  These variables were not mutually exclusive: for example, one class could have both project groups and graded groups.

Another course structure variable was **synchronous**, which encoded whether students primarily participated in lecture synchronously, whether that be online or in-person. We chose to focus on the mode of student participation rather than class delivery for this factor, because most of the courses included in the study had synchronous lectures, even during remote and hybrid learning. The class delivery was also captured in the variable **in-person**, an indicator of whether the mode of instruction was predominantly online or primarily in-person.  This variable distinguished the 2020 – 2021 academic year from Fall 2021. Lastly, in the Fall 2021 semester, variables about the course size and whether final course grades were determined based on a class-wide curve were also considered.

*Student identity variables*

Multiple variables about student identity were considered for the models. These included race/ethnicity, sex, and year in school. **Race/ethnicity** were defined using the following criteria: a student was classified as *Hispanic* if they selected Hispanic in the survey, regardless of whether they selected a race. Among those who identified as non-Hispanic, race was categorized as *White, Black*, *Asian*, *Other (*Middle Eastern, Pacific Islander, preferred not to give a response, or gave no response), and *Multi-racial* if two or more races were selected. The multi-racial category was combined with "Other" to protect the identities of the students due to the small sample size in this group across all terms.

**Results**

*Descriptive Statistics*

We begin by examining the identity factors for students who completed the survey during the study period.

*Table 11: Demographics of survey respondents*

| Factor | **Fall '20** (N = 229) | **Spring '21** (N = 122) | **Fall '21** (N = 128) |
|---|---|---|---|
| **Race** | | | |
| Hispanic | 22 (9.6%) | 14 (11%) | 20 (16%) |
| White non-Hispanic | 103 (45%) | 53 (43%) | 45 (35%) |
| Black non-Hispanic | 23 (10%) | 8 (6.6%) | 8 (6.2%) |
| Asian non-Hispanic | 59 (26%) | 35 (29%) | 35 (27%) |
| Other non-Hispanic | 22 (9.6%) | 12 (9.8%) | 20 (16%) |
| **Sex** | | | |
| Female | 141 (62%) | 77 (63%) | 92 (72%) |
| Male | 88 (38%) | 45 (37%) | 36 (28%) |
| **Year** | | | |
| First-year | 145 (63%) | 71 (58%) | 95 (74%) |
| Second – Fourth year | 84 (37%) | 51 (42%) | 33 (26%) |

Table 1 shows the distribution of demographics for the survey respondents in all three semesters of quantitative data collection. The observed differences in the distributions of the demographic variables partially reflect semester-to-semester differences in the survey responses. Demographic information for all students in participating introductory courses was not available for our study, but most trends are consistent across semesters and align with observations from other courses and programs at Duke.

The majority of students in the data identify as White non-Hispanic or Hispanic, while the fewest percentage of students identify as Black non-Hispanic or Other. A majority of students in each semester identify as female, with a larger percentage of female students in the Fall 2021 survey. A majority of the students are first-years, reflecting the introductory level of these courses.

*Structural equation model*

The final model is shown in Table 2.

*Table 22: Statistical model of connectedness and learning*

| Latent variable | Predictor variable | Estimate (SE) |
|---|---|---|
| Learning support | Connectedness | 0.50 (0.04)*** |
| Connectedness | Hispanic | -0.05 (0.12) |
| | Black non-Hispanic | -0.45 (0.15)*** |
| | Asian non-Hispanic | -0.14 (0.09) |
| | Other non-Hispanic | -0.27 (0.13)* |
| | First-year | 0.05 (0.09) |
| | Male | 0.003 (0.08) |
| | Synchronous | 0.37 (0.09)*** |
| | Project Groups | 0.11 (0.10) |
| | Spring2021 | 0.04 (0.10) |
| | Fall2021 | 0.15 (0.10) |

\* $p < 0.05$; \*\* $p < 0.01$; \*\*\* $p < 0.001$

Based on the structural equation model, there are three factors related to student identity and behavior that have a statistically significant association with connectedness. Compared to White non-Hispanic students, the average connectedness score for Black non-Hispanic is about 0.45 points lower, and the average connectedness score for Other non-Hispanic students is about 0.27 points lower, holding the other factors constant. (The CCS-SF is scored out of 32 points, with each latent factor, learning support and connectedness, scored out of 16 points, such that higher scores indicate stronger sense of classroom community (Cho & Demmans Epp, 2019).) Students who generally participated in in-person, remote, or hybrid class sessions synchronously have an average connectedness score about 0.37 points higher compared to students who did not have regular synchronous participation. It is notable that after preliminary modeling and in this final model, none of the classroom structure variables were statistically significant after accounting for the student-specific variables. Lastly, we observe that for each point in the student's sense of connectedness, their sense of being in a supportive learning environment increases by 0.5 points, on average. This indicates that students who feel connected to others in the course also perceive the learning environment as being welcoming and supportive, where there are common goals and expectations. These results are discussed further in the Discussion and Conclusion.

## Qualitative Analyses

### Methodology

*Instrument*

Following initial quantitative data collection during the 2020-21 academic year, the authors began the process of investigating methods for gathering and analyzing qualitative data. Discussions surrounding the value of qualitative research methodologies focused on asking questions related to *why* and *in what ways* students feel or do not feel a strong sense of classroom community. The quantitative data established that students' feelings of connectedness and learning support varied based on their identities and the characteristics of their courses, but the researchers were interested in learning the context of these responses.

The authors discussed whether key informant interviews, focus groups, or a mix of the two would be most beneficial to the goals of the research. Given the challenges of recruitment during any semester, but particularly during Fall 2021 amid the ongoing restrictions due to the Covid-19 pandemic, it was decided that one mode of data collection would be feasible and sufficient. The authors ultimately chose to conduct focus groups, as these are unique in their ability to capture group opinions in a social context, and thus are particularly useful in educational settings.

*Protocol Design*

The focus group protocol was developed after consulting with the literature on the use of focus groups for student-focused research in education (e.g., Wright & Hendershott, 1992) with special attention paid to studies that included include themes of equity and/or sense of belonging (e.g., Ching & Roberts, 2020). Particular efforts were made to mitigate the bias of the researchers. Open-ended questions and activities allowed focus groups participants to guide the conversation while staying on the topic of belonging in introductory STEM courses. The focus groups were intentionally designed to include between 3-5 participants, which would be enough to hold a discussion but could be easily moderated. Table tents were provided at the beginning of the focus group discussion and participants wrote their first names only. Participants were notified that the discussion would be recorded and their identities would remain confidential. Each focus group discussion lasted around one hour.

Each focus group began with an introduction, the reading and signing of consent forms, and the establishment of ground rules for the group. The overarching topics covered in the focus group discussions, drawn from the literature on sense of classroom community, were (1) support in class; (2) peer connection; and (3) self-confidence.

*Population*

There were a total of 5 focus groups conducted in late Fall 2021 and early Spring 2022 with 11 total students who took an introductory mathematics or statistics course in Fall 2021. Out of these, 7 were first-year students while 4 were second-years, with ages ranging from 18 to 20. Across all groups, we spoke to 7 females and 4 males, out of which 5 were Asian/American, 6 were White/Caucasian, 1 student was Southeast Asian and 1 was Black/African American. While the focus groups were designed to have 3-5 participants each, due to last-minute cancellations, some focus groups had only 2 students and one had a single student.

*Method of Analysis: Tree Plots*

Focus groups were audio-recorded. The audio-recordings were then transcribed using the transcription service Rev.com. Transcripts were uploaded to NVivo, where they were coded thematically by-hand. Themes were then visualized using hierarchical tree plots. The relative sizes of the rectangles in a plot indicate the relative frequency of the themes in focus group discussion. Figure 1 displays the highest level of the tree plot, and Figures 2 and 3 display the sub-tree-plots corresponding to "Discouraging Community" and "Encouraging Community", respectively.

*Figure 1: Common themes from all focus groups*

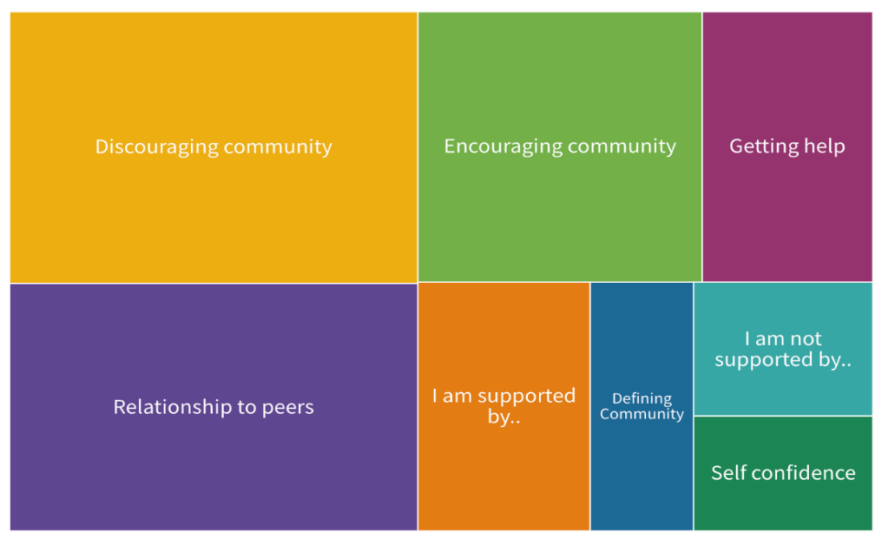

*Figure 1: Common themes from discussion about factors that discourage community*

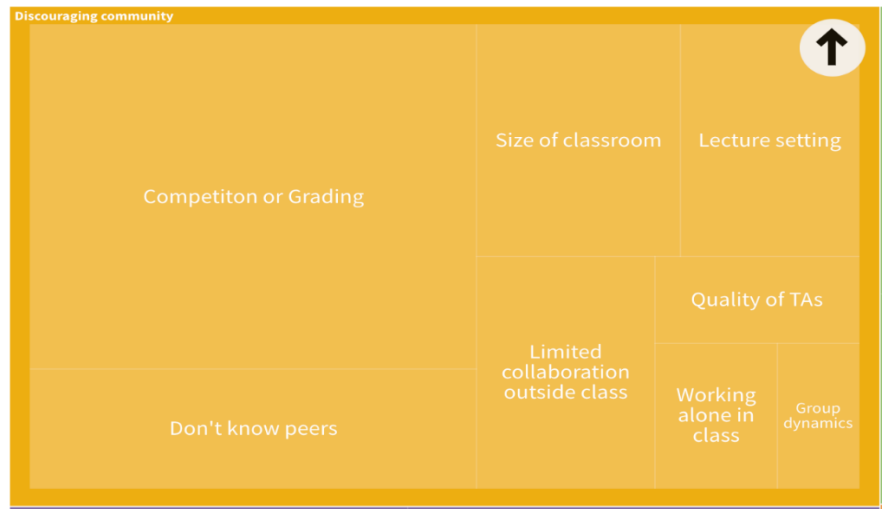

*Figure 2: Common themes from discussion about factors that encourage community*

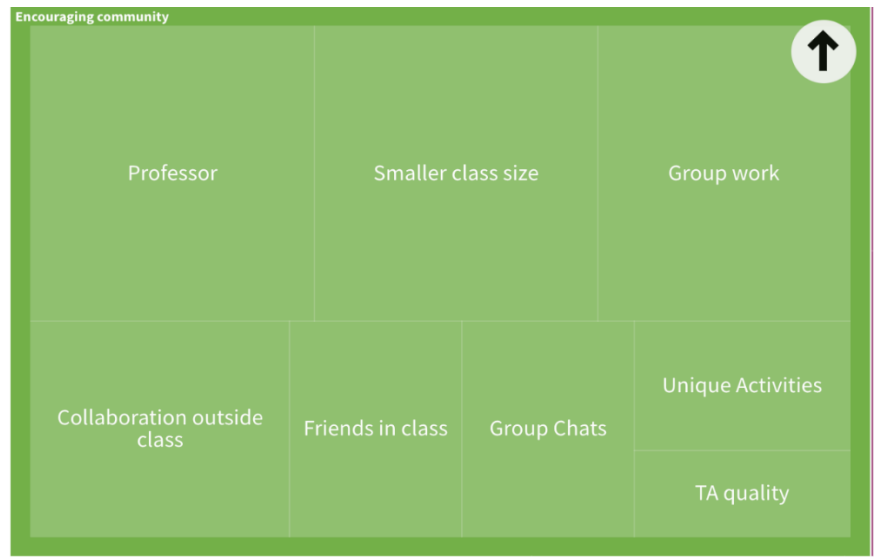

*Word Cloud Methodology*

Word clouds were generated using a method called TF-IDF (term frequency-inverse document frequency), which scores frequently appearing terms in all conversations based on how relevant they were to a specific topic. This was done by multiplying two metrics: how many times a term appears in the context of a focus group topic, and the inverse term frequency of the word across a set of topics. For example, if "class-size" and "professor" both came up 10 times while discussing factors affecting community in a classroom, but class-size came up 100 times across all conversations while professor came up 500 times, the term class-size would be considered more relevant since its relative frequency for this focus group topic is higher.

The formula used for calculating weights (w) based on TF-IDF is $W_{xy} = tf_{x,y} \cdot \log\left(\frac{N}{df_x}\right)$, where $tf_{x,y}$ the frequency of term *x* in topic *y*, $df_x$ is the number of topics in the focus group transcripts where *x* is mentioned, and *N* = total number of topics in the focus group transcripts. Words in a word cloud are sized according to their weights.

Instead of traditional word clouds which size terms by frequency of appearance, our use of TF-IDF ensures that term size increases proportionally to the number of times a term appears in a focus group topic but is offset by the number of topics that contain the term. We considered this essential to our study, because students often mentioned terms in conversations that were generally related to the overarching theme of being a student in a STEM class and were not specifically related to the focus group topic. Before plotting the word clouds, we also removed stop words, which are a set of commonly used words in English – "a", "the", "and", "of", "on", so on.

Once focus group themes were coded, comments for each theme were coded into "positive" and "negative" sentiments by assigning values of +1 or -1.[1] Additionally, comments were also coded with a "neutral" (0) sentiment. An average of these scores makes up the overall sentiment score for a comment, and comment scores are aggregated to make up sentiment scores for a theme.

---

[1] Although the researchers made attempts to use text-based sentiment analyses, we found that there was a significant loss of information due to subtexts/contexts and decided to use a qualitative analysis instead.

Figure 4 displays the word cloud for the theme of grading and sense of community, and Figure 5 displays the word cloud for the theme of personal identity and sense of community.

*Figure 3: Word cloud of common words and phrases about grading and its impact on sense of community, organized by sentiment*

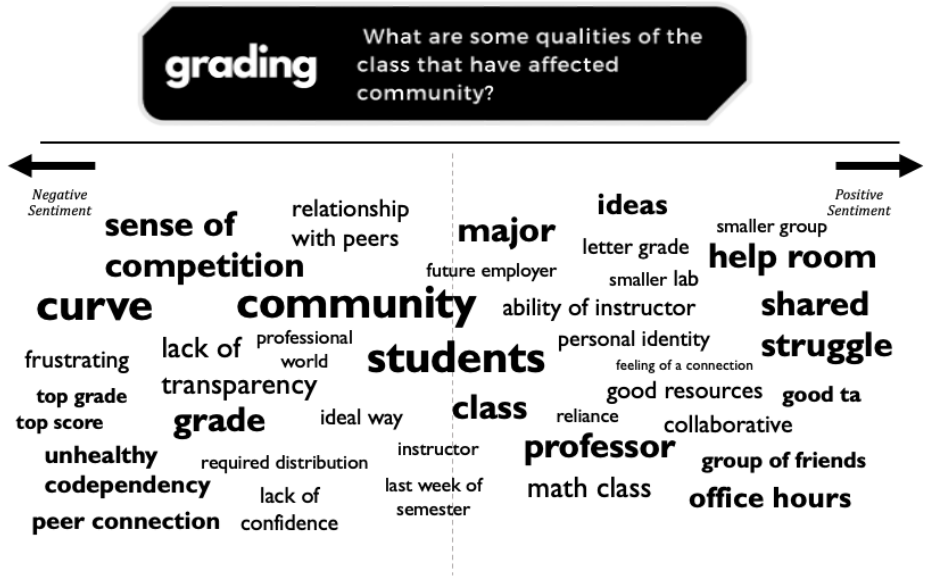

*Figure 4 Word cloud of common words and phrases about personal identity and its impact on sense of community, organized by sentiment*

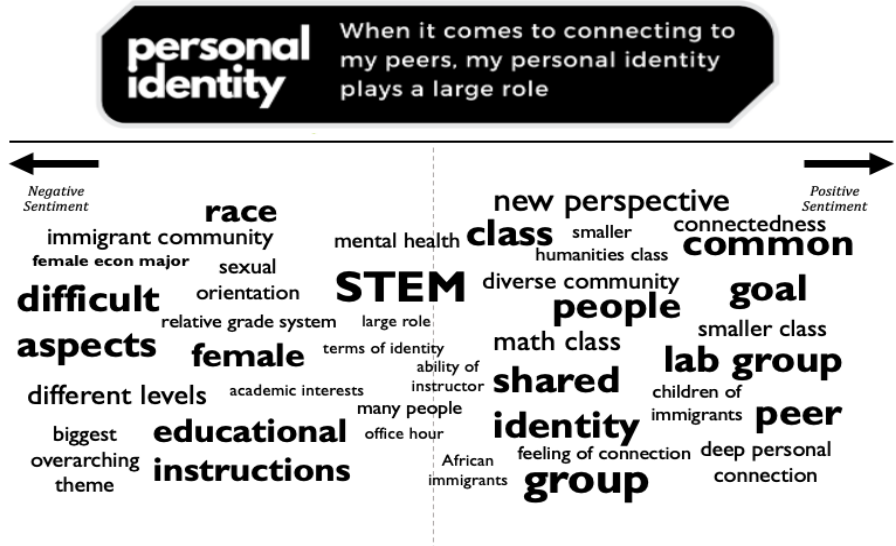

## Results

A few key themes emerged from the analysis of the focus group data, including impact of personal identity and the effect of grading policies and peer-to-peer competition on student sense of community. We examine these themes in closer detail below.

*Personal identity*

In the second topic of the focus group, peer connection, students were first asked the question, "*What are three words you would use to describe your relationship to peers in your class? Take a moment to think*." After a probe, they were then posed with the following to delve into the mediating effects of their identity on their relationships to their peers: "Discuss the following statement: *When it comes to connecting to my peers, my personal identity plays a large role*".

Students were split on whether personal identity plays a role in their building relationships within a class. Notably, for the students for whom personal identity matters, it matters a great deal, as evidenced by the following quotes:

> "*I am with people who I know who understand what I'm going through. So people do matter a lot.*"

> "*I think there's this baseline level of willing to work with someone and working with someone. But it's elevated a lot higher when you share these characteristics that you feel strongly about.*"

Female students spoke about finding it hard to work in all-male groups:

> *"I've been in situations in lab groups or team projects outside of this class specifically where I'm just surrounded by men who will listen to each other, but not me."*

> *"They [male peers] would probably give you a quick answer, but it makes you seem like you were an annoyance."*

> *"You get the feeling that you're being iced slowly."*

As a result, female students find it easier to find community with other women:

> *"I think those experiences have necessarily made me instinctually maybe reach out to women first or hang out with people that I know will actually listen to me."*

International students also spoke about their nationality shaping their identity, even if they didn't prefer it to be that way:

> *"...the moment I start talking to people, 'Oh, which country are you from?'"*
> *"Everyone wants to have their own identity apart from being foreign ."*

> *"When I first came here, I was very lonely because everybody here had everything set. Their parents were like a flight away to our car drive away and my parents were half across the world. So, it probably wasn't just me, it was every foreign international kid who lives here, but for me, my experience felt unique because I was experiencing it firsthand."*

This was echoed in sentiments by children of immigrants:

> *"Because even though I do say being female plays a part in my lived experience, daughter of Asian immigrants is a lot more specific lived experience compared to that. All of us have that shared."*

Lastly, students spoke about race being something they could connect on, especially if they were from marginalized groups in STEM:

> *"...and I relate better to people that I see myself in. I think it's natural."*

Students also recognized the lack of representation in the classroom and expressed a desire to be in more diverse classroom environments:

> *"I'm not really accustomed to seeing other black people in any of my courses."*

> *"I would love to see more representation in some of my lectures, just all my STEM classes, not even just in terms of identity, but experiences and background, that sort of thing."*

*Grading and peer-to-peer competition*

While none of the focus group questions asked about grading directly, the topic nevertheless arose frequently in discussion. Some students disapproved of grading curves because they make students focus on their grades instead of their learning outcomes. These students indicated they would rather be graded more "objectively" for their work instead of being graded relative to their peers:

> *"I disagree with curves because curving means the instructor is compromising on the learning objectives and it reflects that even the instructor does not have* *the students' interest at heart. So what would be a better way is to establish the learning objectives transparently from the very beginning of the course, then structure the course in a way that makes you actually achieve the learning outcomes, instead of using the scores as a reflection of the student."*

Some students talked about an impact of grading curves – the competition they create can discourage students from helping their peers:

*"When there's a class with a humongous curve on all the mid-terms, because then if you help the other students in your class do better, then your grade goes down."*

*"We all support each other, but people are hesitant to reach out, and I think that's because it's slightly competitive because you are being graded against each other with a curve in everything of the class."*

*"...but it can be really problematic, especially when professors are really clear that the top score dictates how it was curved at the beginning of class. Then, people know. They're looking out for who's going to mess up the curve."*

Overall competition was noted to disturb class dynamics. Students said that they were struggling to keep up with classmates:

*"Here's two or three people who are competitive and that just tenses up the whole classroom."*

> *"It's like floating duck syndrome where everyone wants to act like they've got it all together when they're just shuffling."*

This dynamic is made harder by the fact that higher level students seem to want to work together, and others are left to work by themselves:

> *"People who are in higher levels in our class, they like to work by themselves. They don't like to share their answers."*

> *"I think especially because like in high school, we all did mostly have things together. Here it's not the same, it's much harder."*

## Discussion and Conclusion

To answer the research question "*How is undergraduate students' sense of classroom community in introductory mathematics and statistics affected by course attributes and student identities?*", we used a mixed-methods approach that included surveys and a series of focus groups. The survey results allowed us to understand the student experience more broadly and analyze differences between students with different identities and modes of engagement with these courses, while the focus groups allowed us to contextualize the observations from the survey data and more fully understand the experiences of small groups of students.

The results from the analyses showed that personal identity along with how students regularly engage with fellow classmates and the instructor impact their sense of connectedness and perception of the course learning environment. As this project ultimately aims to provide guidance to instructors on identifying students who may currently be feeling left out of classroom communities, and on implementing practices and policies that improve the experiences for these students and the class overall, we provide a few preliminary ideas below. While the Covid-19 pandemic has passed, remote and hybrid learning are very much here to stay, and the lingering effects from the pandemic, including isolation and exacerbated systemic inequality, have only increased the importance of attending to students' sense of belonging and learning support.

**Preliminary implications for classroom practices and policies**

*Foster a supportive learning environment*

The quantitative analysis shows that sense of connectedness is strongly and positively associated with perception of being in a supportive learning environment. This conclusion is supported by the qualitative analysis of the focus group data, as well as the literature (see, e.g., Vavala et al. 2010). Thus, instructors should strive to develop course structures and policies and implement classroom practices that help students feel they have ample opportunity to learn and be successful in the course. Further research is required to identify specific policies and practices that provide this sense, but the focus group data analysis does identify norm-referenced grading as a practice to avoid.

*Be attentive to student identities*

It is important for instructors to understand the experiences of students of different identities and their varying senses of connectedness to their peers and instructor. The quantitative data showed that students from marginalized groups in STEM (in particular, Black, non-Hispanic, Middle Eastern, Pacific Islander, and multi-racial) tend to feel a weaker sense of connectedness than their White non-Hispanic peers. The survey results alone don't indicate why students from these groups feel less connected; however, the focus group data provide some insight: students tend to feel more connected to those with whom they share common identity features, and these commonalities can be harder to find as a member of a marginalized group. While the demographic make-up of an institution or even a single course is subject to forces beyond an instructor's control, there may be other mitigating actions that can be taken to foster a more inclusive classroom space, and attentiveness to diversity of identities and experiences is a place to begin. Further research is required to better understand students' specific experiences and the intersectional aspects of their identities with their perceptions of classroom connectedness.

*Encourage synchronous participation*

Another conclusion from the analysis of survey results is the positive association between synchronous participation and sense of connectedness. This indicates that instructors might foster a stronger sense of belonging by designing their courses in a way that encourages students to attend and participate synchronously. Some focus group participants remarked how they appreciate the times instructors asked them to get to know their peers next to them or did some hands-on activity that felt unique to that class. Thus, we recommend that instructors make the in-class environment engaging to encourage synchronous participation, rather than trying to encourage it through more passive, and potentially punitive means, such as participation points or pop quizzes. Incorporating a single active learning activity such as *think-pair-share*, *peer instruction*, *in-class group work*, or *one minute paper* into a class session is a next step instructors can take to foster more participation and engagement during synchronous sessions.

We emphasize that these suggestions are a starting point only. In addition to identifying limitations of our work and needed areas for further research, we also point instructors to practitioner-focused resources for inclusive teaching practices like Cardon & Womack (2022) and Sathy & Hogan (2022).

**Limitations**

The primary limitation of this study is the sample population: all data were gathered from a single institution. Due to the lack of variation in the way introductory mathematics and statistics courses are taught at this one university, it was challenging for the quantitative model to identify significant associations between different classroom practices and student sense of connectedness and learning support. The nature of the institution from which the data was gathered is also salient: student experiences at a private university in the United States may differ significantly from those at other undergraduate institutions. Likewise, experiences of students of color at a predominantly White institution (PWI) may differ significantly from those at a historically black college or university (HBCU) or Hispanic-serving institution (HSI). Further work is required to investigate these differences.

Another limitation of this study is the variation in data collection methods. Instructors administered the student survey in their own courses, leading to potential inconsistencies in administration across course sections. For example, some instructors may have simply emailed the survey link to students, while others may have provided time in class for completion, affecting response rates. On the instructor data side, there was some ambiguity in terminology and inconsistency in the way data was gathered, both via survey and interview. For example, the authors suspect that group-based learning is significant, as seen in the literature (see, e.g., Parrish et al, 2021), but there is some noise in the definitions of work groups and project groups.

Finally, it is important to note that the relationships found in the quantitative models are not causal. This means, for example, that just as much as students who participated synchronously felt a stronger sense of connectedness, it can be said that students with a stronger sense of connectedness wanted to participate synchronously. Thus, while the model indicates that encouraging synchronous participation may foster classroom community, the direction of the relationship cannot be definitively identified from the existing data.

**Future work**

One of the primary benefits of a mixed-methods approach is the way in which each method gives rise to research questions for the other. Further quantitative study is needed into some of the key themes identified in the focus group discussions, such as grading policies and competition as well as personal identity and the demographics of working groups. A clear next step is to refine the instructor survey to inquire more specifically about the course-level variables of interest. Another possibility would be to add questions to the student survey about their perception of various specific course practices and policies. Most importantly, it is essential that data be gathered from a wider variety of sources and institutions. This will introduce more variability in classroom attributes, and thereby strengthen the ability of the quantitative model to discern which of these practices and policies are significant predictors of student sense of community. Furthermore, gathering data from a variety of institutions, including HBCUs and HSIs, will provide further insight into variations in student experiences in different settings, as mediated by student demographics. Thus, we are currently working to expand this research to include more institutions that together serve a more diverse range of student populations.

Complementarily, additional focus groups would be valuable to probe more deeply and specifically into identity issues and responses to specific classroom practices. One possibility would be to conduct identity-specific focus groups in which students from marginalized groups may feel more comfortable sharing their experiences among peers with the same characteristics.

Indeed, by chance one of the focus groups conducted had such a shared demographic bond, and it gave rise to a particularly poignant discussion. It is clear from the analyses that personal identity is tied to connectedness and belonging, and from the literature that sense of belonging is particularly salient for the STEM persistence of students from marginalized groups like women and students of color (Seymour et al., 2019). Future work is essential to better understand what introductory STEM instructors can do, in addition to long-term efforts to increase diversity of representation, to improve their students' sense of belonging.

**Acknowledgements**

This project was partially supported by the Carry the Innovation Forward Grant from Duke Learning Innovation. Portions of these findings were presented at the International Conference on Learning in July 2024, at the Electronic Conference On Teaching Statistics (eCOTS) on May 24, 2022, and at the Duke-sponsored Teaching Diverse Learners panel on November 2, 2021. We have no conflicts of interest to disclose. We thank Margaret Poulos for her contribution to developing and piloting the focus group protocol, thank Kim Manturuk for her contributions facilitating focus groups, interviewing instructors, and de-identifying data, and thank Andrea Novicki for her contributions in launching the project and conducting instructor interviews.

Burch, M., Lohmann, S., Beck F., Rodriguez N., Di Silvestro, L. & Weiskopf, D. "RadCloud: Visualizing Multiple Texts with Merged Word Clouds," *2014 18th International Conference on Information Visualisation*, 2014, pp. 108-113, doi: 10.1109/IV.2014.72.

Cardon, L.S., & Womack, A.-M. (2022). Inclusive College Classrooms: Teaching Methods for Diverse Learners (1st ed.). Routledge. https://doi.org/10.4324/9781003121633

Cho, J. & Demmans Epp, C. (2019). Improving the Classroom Community Scale: Toward a Short-Form of the CCS. *Presented at the American Educational Research Association (AERA) Annual Meeting,* Toronto, Canada.

Ching, C. D. & Roberts, M. (2020). Equity-minded inquiry series: Conducting student interviews and focus groups. Rossier School of Education, University of Southern California.

Creswell, J. W. (2009). *Research design: Qualitative, quantitative, and mixed methods approaches* (3rd ed). Sage Publications.

Crisp, B. (2010). Belonging, connectedness and social exclusion. *Journal of Social Inclusion*, *1*. https://doi.org/10.36251/josi.14

Dawadi, S., Shrestha, S., & Giri, R. A. (2021). Mixed-Methods Research: A Discussion on its Types, Challenges, and Criticisms. *Journal of Practical Studies in Education*, 2(2), 25–36. https://doi.org/10.46809/jpse.v2i2.20

Dellasega, S. (2021). *Technology-enhanced interaction, residency requirements, and student characteristics in fully online programs and their relationship with student connectedness.* [PhD thesis]. Lindenwood University.

Dawson, S. (2008). A study of the relationship between student social networks and sense of community. *Educational Technology and Society*, 11 , 224-238.
28

Rovai, A. (2002a). Development of an instrument to measure classroom community. *The Internet and Higher Education*, *5*(3), 197–211. https://doi.org/10.1016/S1096-7516(02)00102-1

Rovai, A. (2002b). Sense of community, perceived cognitive learning, and persistence in asynchronous learning networks. *The Internet and Higher Education*, *5*(4), 319–332. https://doi.org/10.1016/S1096-7516(02)00130-6

Rovai, A. & Jordan, H. (2004). Blended Learning and Sense of Community: A Comparative Analysis with Traditional and Fully Online Graduate Courses. *The International Review of Research in Open and Distributed Learning*, *5*(2). https://doi.org/10.19173/irrodl.v5i2.192

Sathy, V., & Hogan, K.A. (2022). *Inclusive Teaching*: *Strategies for Promoting Equity in the College Classroom*. Morgantown: West Virginia University Press.

Sappleton, N. & Lourenço, F. (2016). Email subject lines and response rates to invitations to participate in a web survey and a face-to-face interview: The sound of silence. *International Journal of Social Research Methodology*, *19*(5), 611–622.

Seymour, E. (1997). *Talking About Leaving*. Westview Press.

Seymour, E., Hunter, A.-B., Thiry, H., Weston, T. J., Harper, R. P., Holland, D. G., Koch, A. K. & Drake, B. M. (2019). *Talking about Leaving Revisited Persistence, Relocation, and Loss in Undergraduate STEM Education*.

Summers, J. J., & Svinicki, M. D. (2007). Investigating classroom community in higher education. *Learning and Individual Differences*, *17*(1), 55–67. https://doi.org/10.1016/j.lindif.2007.01.00631

**Corresponding Author**

Shira Viel, Department of Mathematics, Duke University, 120 Science Drive, 117 Physics Building, Campus Box 90320, Durham, NC 27708, United States. Email: mailto:shira.viel@duke.edu. ORCID: https://orcid.org/0000-0003-0613-7596.